\documentclass[%
 aip,
 amsmath,amssymb,
 reprint,%
]{revtex4-1}

\usepackage{graphicx}
\usepackage{dcolumn}
\usepackage{bm}

\usepackage[utf8]{inputenc}
\usepackage[T1]{fontenc}
\usepackage{mathptmx}

\usepackage{hyperref}
\hypersetup{colorlinks = true, allcolors = blue, pdfauthor={Pavel Kuptsov}}

\newcommand{\fighere}[2]{\centering{
    \includegraphics[width=#1\columnwidth]{#2}}}

\newcommand{\figherewide}[2]{\centering{
    \includegraphics[width=#1\textwidth]{#2}}}

\newcommand{\dimky}{D_{\text{KY}}}

\newcommand{\Covar}{\operatorname{Covar}}
\newcommand{\const}{\mathrm{const}}

\begin{document}

\preprint{AIP/123-QED}

\title[Hyperbolic hyperchaos]{Route to hyperbolic hyperchaos in a
  nonautonomous time-delay system}

\author{Pavel V. Kuptsov}

  \email{kupav@mail.ru}

 \affiliation{Laboratory of topological methods in dynamics,
   National Research University Higher School of Economics,
   Nizhny Novgorod, 25/12 Bolshay Pecherskaya str., Nizhny
   Novgorod 603155, Russia}
 
 \affiliation{Institute of Electronic Engineering and Instrumentation, Yuri
   Gagarin State Technical University of Saratov, Politekhnicheskaya
   77, Saratov 410054, Russia}

 \affiliation{Kotel’nikov’s Institute of Radio-Engineering and
   Electronics of RAS, Saratov Branch, Zelenaya 38, Saratov 410019,
   Russia}

\author{\fbox{Sergey P. Kuznetsov}\,}
\affiliation{Kotel’nikov’s Institute of Radio-Engineering and
  Electronics of RAS, Saratov Branch, Zelenaya 38, Saratov 410019,
  Russia}

\date{\today} 

\begin{abstract}
  We consider a self-oscillator whose excitation parameter
  is varied. Frequency of the variation is much smaller then
  the natural frequency of the oscillator so that
  oscillations in the system are periodically excited and
  decay. Also a time delay as added such that when the
  oscillations start to grow at a new excitation stage they
  are influenced via the delay line by the oscillations at
  the penultimate excitation stage. Due to a nonlinearity
  the seeding from the past arrives with a doubled phase so
  that oscillation phase changes from stage to stage
  according to chaotic Bernoulli-type map.  As a result, the
  system operates as two coupled hyperbolic chaotic
  subsystems. Varying the relation between the delay time
  and the excitation period we affect a coupling strength
  between these subsystems as well as intensity of the phase
  doubling mechanism responsible for the hyperbolicity. Due
  to this, a transition from non-hyperbolic to hyperbolic
  hyperchaos occurs. The following steps of the transition
  scenario are revealed and analyzed: (a) an intermittency
  as alternation of long staying near a fixed point at the
  origin and short chaotic bursts; (b) chaotic oscillations
  with often visits to the fixed point; (c) plain hyperchaos
  without hyperbolicity after termination visiting the fixed
  point; (d) transformation of hyperchaos to hyperbolic
  form.
\end{abstract}


\maketitle

\begin{quotation}
  A hyperchaotic attractor has at least two positive
  Lyapunov exponents, i.e., unlike simple chaotic attractor
  its phase space contains two or more expanding
  directions. In applications employing deterministic chaos
  hyperchaotic systems are usually more preferable since
  their dynamics is more complicated in comparison with mere
  chaotic systems. However many chaotic as well as
  hyperchaotic systems have actually a quasiattractor, i.e.,
  a limit set containing stable periodic orbits, so that
  their dynamics is not so stochastic as expected. Good
  scholastic properties justified in rigorous mathematical
  sense are guarantied for so called hyperbolic
  attractors. Systems with attractors of this type
  demonstrate strong and structurally stable chaos that is
  insensitive to variation of functions and parameters in
  the dynamical equations, to noises, interferences etc.
  Our study in the present paper will be focused on a
  nonautonomous time-delay system with a hyperbolic
  hyperchaotic attractor. This system operates as two
  coupled hyperbolic chaotic subsystems. Varying its
  parameters we can control a coupling strength between
  these subsystems as well as a mechanism responsible for
  their hyperbolic chaos. Due to this, a transition from
  non-hyperbolic to hyperbolic hyperchaos occurs. The
  following steps of the transition scenario are revealed
  and analyzed: (a) an intermittency as alternation of
  staying near a fixed point and chaotic bursts; (b)
  wandering between the fixed point and chaotic subset that
  appears near it; (c) plain hyperchaos without
  hyperbolicity after termination visiting the fixed point;
  (d) transformation of hyperchaos to hyperbolic form.
\end{quotation}

\section{Introduction}

Attractors characterized by two or more positive Lyapunov
exponents are called hyperchaotic. The simplest and trivial
example is provided by several chaotic systems with weak
coupling whose attractor is mere a direct sum of partial
attractors~\cite{KapChua94}. However the phase space
dimension in this case is superfluous. The smallest possible
dimension is four: two expanding directions, one contracting
and one neutral. The first nontrivial hyperchaotic attractor
in a four-dimensional system was proposed by
R\"{o}ssler~\cite{Rossler1979}.

A lot of studies were done around hyperchaotic dynamics
since that time and it still attracts an interest of
researchers. In a recent paper~\cite{stankevich2018chaos}
routes of transition to hyperchaotic dynamics associated
with different bifurcations of periodic and quasi-periodic
regimes are revealed for coupled antiphase driven Toda
oscillators. Paper~\cite{garashchuk2019hyperchaos} reports a
study of two coupled contrast agents being a micrometer size
gas bubbles encapsulated into a viscoelastic shell. Such
bubbles are used for enhancing ultrasound visualization of
blood flow and have other promising applications like
targeted drug delivery and noninvasive therapy. For the
onset of hyperchaotic dynamics in this system a new
bifurcation scenario is proposed that includes an appearance
of a homoclinic chaotic attractor containing a saddle-focus
periodic orbit with its two-dimensional unstable
manifold. Moreover it is shown that hypechaotic attractors
are stable with respect to perturbations that destroy
synchronization manifold in the considered system.
Radiophysical experiments with hyperchaotic dynamics as well
as corresponding theoretical analysis are reported in
Ref.~\cite{stankevich2019chaos}. It is shown that as a
result of a secondary Neimark–Sacker bifurcation, a
hyperchaos with two positive Lyapunov exponents can occur in
the system. A comparative analysis of chaotic attractors
born as a result of loss of smoothness of an invariant
curve, as a result of period-doubling bifurcations, and as a
result of secondary Neimark–Sacker bifurcation is carried
out.

Hyperchaotic systems have more then one expanding directions
in phase space so that their dynamics is more complicated in
comparison with mere chaotic systems. In particular the
prediction time of hyperchaotic regimes can be is much less
than that for chaos~\cite{Nikolov06}. Thus, hyperchaotic
oscillators are employed when the complexity of a signal is
crucial, for example for secure
communications~\cite{KocarevParlitz95, PengDing96, Yaowen00,
  WangWenxin19} and for image
encryption~\cite{Sun08,Zhu12,Cryptanalysis1,Cryptanalysis2}.
One more promising application of hyperchaotic systems is
damage assessment based on using a steady-state chaotic
excitation~\cite{Torkamani2012}.

For applications where complexity is critical one must take
into account that many chaotic systems have actually a
quasiattractor, i.e., a limit set containing stable periodic
orbits, so the observable dynamics can be not so irregular
as expected and be dramatically sensitive to small
variations of parameters. Good scholastic properties
justified in rigorous mathematical sense are guarantied for
so called hyperbolic attractors. Systems with attractors of
this type, like, for example, the Smale-Williams solenoid,
demonstrate strong and structurally stable chaos that is
insensitive to variation of functions and parameters in the
dynamical equations, to noises, interferences
etc.~\cite{HyperBook12}.

Hyperbolic attractors are composed exclusively of saddle
trajectories~\cite{Smale67,Anosov95,KatHas95}. For all their
points a space of small perturbations (tangent space) is
split into a direct sum of everywhere exponentially
expanding and contracting subspaces. In the phase space
these subspaces are tangent to corresponding expanding and
contracting manifolds. In autonomous flow systems, in
addition, there is a one-dimensional neutral tangent
subspace of perturbations along a trajectory that
corresponds to marginally stable shifts in time. Necessary
and sufficient condition of the hyperbolicity is that
absence of tangencies between stable, unstable and neutral,
if any, manifolds; only intersections at nonzero angles are
admitted.

Due to their great potential importance for applications,
structurally stable chaotic systems with hyperbolic
attractors obviously have to be a subject of priority
interest, like rough systems with regular dynamics in the
classic theory of
oscillations~\cite{AndrPontr,AndrKhaikVitt}. However, for many
years the hyperbolic attractors were commonly regarded only
as purified abstract mathematical images of chaos rather
than something intrinsic to real world systems. A certain
progress in this field has been achieved recently when many
examples of physically realizable systems with hyperbolic
attractor have been purposefully
constructed~\cite{HyperBook12,KuzUFN11}.

A hyperchaotic system can be hyperbolic. Obvious example
consists of two ordinary hyperbolic systems with weak
coupling. Due to their structural stability the
hyperbolicity of the subsystems survives at least when the
coupling is small so that the whole system is hyperbolic and
hyperchaotic. 

An interplay between hyperbolicity and hyperchaos was
studied in Refs.~\cite{HyperSpace09,
  UDVHyp2012}. Paper~\cite{HyperSpace09} reports the
scenario of transition to hyperchaos in a one-dimensional
spatially distributed medium with local hyperbolic
chaos. When its length is small all spatial elements
oscillate synchronously and demonstrate hyperbolic chaos. As
the length grows the second Lyapunov exponent becomes
positive, and spatial homogeneity is destroyed. But the
hyperbolicity survives so that the system demonstrates a
hyperbolic hyperchaos. Further growth of the length results
in the emergence of the third positive Lyapunov exponent
accompanied by violation of the
hyperbolicity.

Paper~\cite{UDVHyp2012} considers the violation of
hyperbolicity and transition to hyperchaos in a chain of
diffusively coupled oscillators with hyperbolic chaos. It is
shown that it occurs via an intermittency and so called
unstable dimension variability (UDV). UDV regime is
characterized by coexistence in the chaotic attractor of
invariant periodic or chaotic orbits with different number
of unstable
directions~\cite{Kostelich1997,Pereira2007}. Since
trajectories of the system can pass close to these orbits,
the dimensions of their unstable and stable manifolds vary.
The UDV and intermittency as a part of scenario of
transition to hyperchaos were also reported in
Refs.~\cite{KapMaist00,Yanchuk01,YanchukSymmetry01}. Other
reveled details of the transition included a blowout
bifurcation and bubbling.

Systems with a time-delay feedback combine simplicity of
implementation and rich complexity of dynamics. Examples of
such systems are wide-spread in electronics, laser physics,
acoustics and other fields~\cite{VihLaf13}. Recently several
examples were suggested as physically realizable devices for
generation of rough hyperbolic
chaos~\cite{KuzPonom08,Autonom10,
  KuzPik08,Baranov10,KuzKuz13,Arzh14}. Though rigorous
mathematical proof of their hyperbolicity is not performed
yet, the hyperbolicity of these systems is confirmed
numerically in Refs.~\cite{HypDelay2016,HypManyDelay2018}.

Our study in the present paper will be focused on a
nonautonomous time-delay system with a hyperbolic attractor
suggested in Ref.~\cite{KuzPonom08}.  As discussed in
paper~\cite{BaranovHyper10}, varying parameters of this
system one can also obtain hyperbolic hyperchaotic
attractors with as many positive Lyapunov exponents as
required. In this paper we study a hyperchaotic attractor
with two positive Lyapunov exponents. We perform a numerical
test that confirms its hyperbolicity and analyze the details
of transition to hyperbolic hyperchaotic regime.

The paper is organized as follows. In Sec.~\ref{sec:sys} we
introduce a system and discuss how it operates. Also we
briefly review studding methods that are used
below. Section~\ref{sec:analysis} discusses the transition
from non-hyperbolic to hyperbolic hyperchaos. It is divided
in several subsections: Subsec.~\ref{subsec:lyap} is focused
on Lyapunov exponents, angle between expanding and
contracting subspaces, and the Kaplan-Yorke dimension;
Subsec.~\ref{subsec:pdf} represents two-dimensional
distributions of various characteristic values on the
attractor; in Subsec.~\ref{subsec:subsets} we deal with the
degenerated invariant subsets of the attractor; and
Subsec.~\ref{subsec:diffus} discusses the large time-scale
behavior of finite time Lyapunov exponents (FTLEs). In
Sec.~\ref{sec:outline} we outline the obtained results.

\section{\label{sec:sys}The system and methods of analysis}

\begin{figure}
  \fighere{0.99}{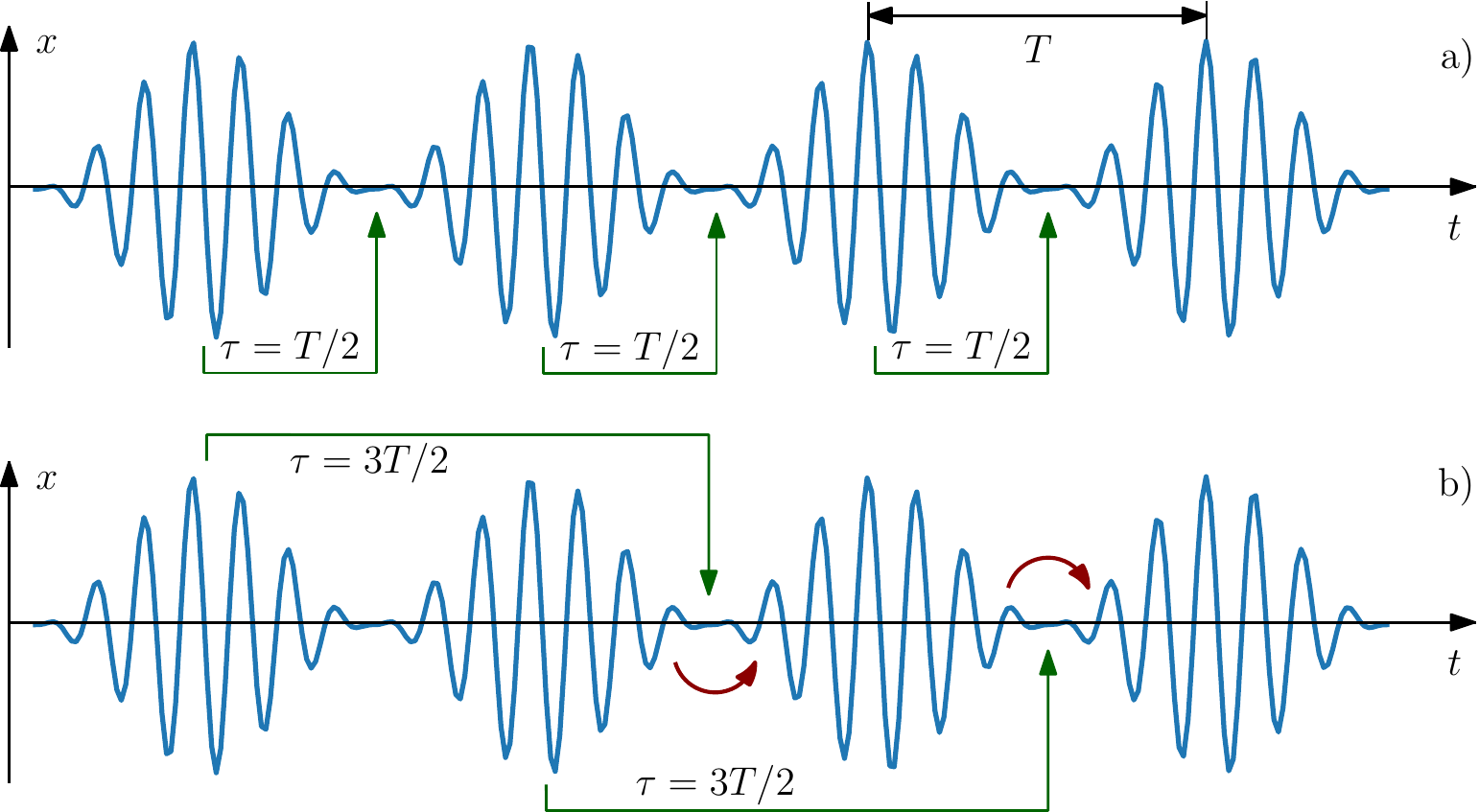}
  \caption{Operation of the system~\eqref{eq:sys}: (a)
    hyperbolic chaos, and (b) hyperbolic hyperchaos with two
    positive Lyapunov exponents. Polyline arrows shows the
    seeding transfer between excitation stages, and arc
    arrows show how the interaction between the subsystems
    occurs.}
  \label{fig:operation}
\end{figure}

We will consider a nonautonomous system based on the van der
Pol oscillator of natural frequency $\omega_0$ supplied with
a specially designed time-delay feedback~\cite{KuzPonom08}:
\begin{equation}
  \label{eq:sys}
  \ddot x - [A\cos(2\pi t/T)-x^2]\dot x +
  \omega_0^2 x = \epsilon x(t-\tau) \dot x(t-\tau)\cos \omega_0 t.
\end{equation}
The parameter controlling the oscillator excitation is
modulated with the period $T$ and amplitude $A$. Modulation
is slow, $T\gg 2\pi/\omega_0$, so that the positive
half-period is sufficiently long for the periodic
oscillations to grow up. Then the oscillations decay and
grow again at the next excitation stage corresponding to the
next positive half-period of the modulation, see
Fig.~\ref{fig:operation}. The main harmonic at $n$th
excitation stage can be written as $\sin(\omega_0t+\phi_n)$
where phase $\phi_n$ is controlled via the delay line.  If
the retarding time $\tau$ is close to $T/2$, as shown in
Fig.~\ref{fig:operation}(a), the emergence of the
self-oscillations at each stage of activity is stimulated by
a signal at the previous activity stage whose dominating
harmonic is $\sin(\omega_0t+\phi_{n-1})$. When it passes
through a nonlinear delayed terms the resonant harmonics
$\sin(\omega_0t+2\phi_{n-1})$ with the doubled phase
appears:
\begin{equation}
  \begin{gathered}
    x(t-\tau) \dot x(t-\tau)\cos \omega_0 t = \\%
    \omega_0 \sin(\omega_0t+\phi_{n-1}) \cos(\omega_0t+\phi_{n-1})\cos \omega_0 t= \\%
    (\omega_0/2) \sin(2\omega_0t+2\phi_{n-1})\cos \omega_0 t = \\%
    (\omega_0/4) \sin(\omega_0t+2\phi_{n-1}) + \ldots
  \end{gathered}
\end{equation}
This harmonic determines the phase $\phi_n$ of the new
excitation stage when oscillations starts to grow. To avoid
its further influence and allow new oscillations to grow
freely parameter $\epsilon$ is taken small. As a
result, we get a sequence of oscillation trains with phases
at successive excitation stages obeying a chaotic
Bernoulli-type map,
\begin{equation}
  \label{eq:bernoul1}
  \phi_{n}=2\phi_{n-1}+\const \mod 2\pi.
\end{equation}
(A constant addition appears since we transfer phase to the
beginning of the stage and measure it in the middle.)
According to argumentation in Ref.~\cite{KuzPonom08}, this
means that the attractor for the Poincar\'e map, that
corresponds to states obtained stroboscopically at
$t_n = nT$, is a Smale-Williams solenoid, and the respective
chaotic dynamics is hyperbolic with the first Lyapunov
exponent close to $\log 2$. In Ref.~\cite{HypDelay2016} this
argumentation is confirmed via numerical test for $\tau$
values between approximately $T/4$ and $3T/4$.

The described mechanism of doubled phase transfer between
excitation stages that results in hyperbolic chaos is
reported for the first time in Ref.~\cite{Kuznetsov05} and
discussed in more detail in Refs.~\cite{KuzUFN11,
  HyperBook12}.

As reported in the paper~\cite{BaranovHyper10}, using longer
retarding times, say $\tau=3T/2$ that provides the seeding
of a new excitation stage from the stage before the previous
one, see Fig.~\ref{fig:operation}(b), it is possible to
observe hyperchaos with two positive Lyapunov exponents. In
this case the map for phases at successive excitation stages
looks as follows:
\begin{equation}
  \label{eq:bernoul2}
  \phi_{n}=2\phi_{n-2}+\const \mod 2\pi.
\end{equation}
The sequence of phases now contains two independent chaotic
sequences whose elements alternate. Thus, the
system~\eqref{eq:sys} in this case can be treated as
consisting of two weakly coupled hyperbolic chaotic
subsystems whose interaction produces hyperchaotic
hyperbolic attractor. The subsystems interact on the
boundary between excitations stages see the arc arrows in
Fig.~\ref{fig:operation}(b), and the hyperbolicity mechanism
brings here the seeding with a doubled phase, see the
polyline arrows. The type of dynamics depends on a relation
between amplitudes of these two channels. This can be
controlled by varying $\tau$ or $T$. If $\tau=3T/2$, as in
Fig.~\ref{fig:operation}(b), the hyperbolicity mechanism has
the highest amplitude and thus dominates the coupling. In
this case the subsystems operate almost independently
producing the hyperbolic hyperchaos. If $\tau\approx T$, the
hyperbolicity channel is the weakest, so that the coupling
prevails. In what follows, decreasing $T$ we will observe
the transition to hyperbolic hyperchaos as a result of the
decrease of the relative coupling strength.

In general, for
\begin{equation}
  \label{eq:hchaos_cond}
  \tau=(k-1/2)T,\;\; k=1,2,3\ldots,
\end{equation}
the system~\eqref{eq:sys} may be expected to have a
hyperchaotic attractor with $k$ positive Lyapunov exponents
equal to $k^{-1}\log 2$ \cite{BaranovHyper10}.

In this paper we will focus on the case $k=2$ for 
\begin{equation}
  \label{eq:params}
  \tau=12, \; A=3,\; \epsilon=0.3, \; \omega_0=2\pi.
\end{equation} For sufficiently large modulation period
$T=10$ dynamics of
Eq.~\eqref{eq:sys} is regular. When $T$ gets smaller,
hyperchaotic attractor appears, then it undergoes certain
transformations, and finally becomes hyperbolic. At $T=8$
the condition~\eqref{eq:hchaos_cond} is fulfilled exactly.

Due to presence of the delay, the system~\eqref{eq:sys} is
infinite-dimensional. Dealing with its computational model,
we introduce discretization along time variable so that the
dimension of the resulting model depends on the number of
steps on the delay interval. Setting the step size
$\Delta t=0.01$ and taking the retarding time $\tau=12$ we
obtain for the second order delay differential
equation~\eqref{eq:sys} a numerical model whose phase space
dimension is $N=2402$.

We will analyze the system~\eqref{eq:sys} numerically using
Lyapunov analysis. In brief, it includes studying of
expanding and contracting properties of perturbation vectors
and volumes spanned by these vectors as the system runs
along a trajectory. The perturbation vectors are assumed to
be infinitely small in magnitude and form a linear tangent
space. The dimension of this space is equal to the phase
space dimension $N$.

Globally, i.e., for an infinitely long trajectory,
properties of the tangent vectors are described by a set of
Lyapunov exponents $\lambda_i$, $i=1,2,\ldots N$, sorted in
descending order. They can be treated in two ways. On the
one hand, the sum of the first $k$ Lyapunov exponents is an
average rate of exponential expansion (or contraction, if
negative) of every typical $k$-dimensional volume in the
tangent space. On the other hand, the $n$th Lyapunov
exponent is an average rate of exponential expansion of the
$n$th covariant Lyapunov vector (CLV). These vectors are
named ``covariant'' since $n$th vector at time $t_1$ is
mapped by a tangent flow to the $n$th vector at time $t_2$
for any $t_1$ and $t_2$. There is a unique set of $N$ such
vectors. An arbitrary tangent vector does not have this
property and merely converges to the first CLV. Two
algorithms for computation of CLVs were first reported in
the pioneering works~\cite{GinCLV,WolfCLV}. See also
paper~\cite{CLV2012} for more detailed explanation and
discussion of one more algorithm. Also see
book~\cite{PikPol16} for a survey.

Using the Lyapunov exponents one can compute Kaplan-Yorke
dimension of the attractor~\cite{KYDim}.
\begin{equation}
  \dimky = m + \frac{\sum_{i=1}^m \lambda_i}{|\lambda_{m+1}|},
\end{equation}
where $m$ is such that $\sum_{i=1}^m \lambda_i>0$ and
$\sum_{i=1}^{m+1} \lambda_i<0$. The Kaplan-Yorke dimension
is related with the information dimension and is an upper
estimate for the Hausdorff dimension of an
attractor~\cite{GrasProc83}.

Local structure of the attractor can be analyzed using
finite time Lyapunov exponents (FTLEs) $\ell_i$. There are
two different sorts of these exponents. One is obtained in
the course of the standard algorithm for Lyapunov exponents
when we iterate a set of tangent vectors and periodically
orthonormalize them using Gram-Schmidt or QR
algorithms. Logarithms of their norms divided by the time
step between the orthonormalizations may be called
Gram-Schmidt FTLEs. This sort of FTLEs characterizes local
volume expanding properties in the tangent space. The sum of
the first $k$ Gram-Schmidt FTLEs is a rate of local
exponential expansion of a typical $k$-dimensional tangent
volume. Their individual values except the first one have no
much sense. Another sort of FTLEs are computed as local
exponential expansion rates for CLVs. They characterize
expansion or contraction for individual vectors in tangent
space. In more detail, the difference between these two
sorts of FTLEs is discussed in Ref.~\cite{PsHyp2018}. In
what follows we will consider the CLV based FTLEs.

Dealing with FTLEs for flow systems, one have to choose an
appropriate time step. This is not so obvious since the
choice must be related somehow with intrinsic attractor time
scales, that are usually a priori unknown. One way to put
the FTLEs analysis on the solid ground is to consider them
on infinitesimally small times. Such instant FTLEs were
introduced in paper~\cite{PsHyp2018}. For discrete time
systems, however, one can compute one step FTLEs. Since the
system under consideration in this paper operates under
external forcing with period $T$, it is natural to consider
the corresponding stroboscopic map for it. Thus all FTLEs
below will be computed for one step of this map, i.e. for
one period $T$ in terms of the original flow system.

Another way of using FTLEs is to consider them on
asymptotically long times. For large time scale the
Gram-Schmidt and the CLV based FTLEs
coincide~\cite{CLV2012}, so it is reasonable to use the
Gram-Schmidt ones since they require much less computational
efforts. Due to the decay of correlations for a typical
chaotic processes on large time scales, pairwise covariances
of Lyapunov sums $L_i$ (FTLEs not divided by time step) are
expected to grow linearly. The matrix $D_{ij}$ of the
corresponding growth rates is introduced and studied in
Ref.~\cite{StatMechLyap11}. Below we analyze the covariances
and show that for some parameter values they demonstrate
power law instead of the expected linear growth.

To characterize the hyperbolicity we will use the angle
criterion. Chaotic attractor is called hyperbolic when all
its trajectories are of saddle type. It means that its
expanding and contracting manifolds never have
tangencies. Verification of this property can be done by
checking the angles between tangent subspaces spanned by
CLVs corresponding to positive and negative Lyapunov
exponents (or, more rigorously, the smallest principal angle
between these subspaces). The angle $\theta_i$ is the angle
between a subspace spanned by the first $i$ CLV and the
subspace spanned by all the rest of them. If a discrete time
system has $k$ positive Lyapunov exponents and all others
are negative then it will be hyperbolic if $\theta_k$ never
vanish along trajectories on the attractor. As we consider a
system with two positive Lyapunov exponents, the indicating
angle is $\theta_2$. Notice that in actual computations
starting from random initial conditions we will never get
exact zero angle. Instead, a typical trajectory can pass
arbitrary close to points with zero angles. Thus verifying
the hyperbolicity we can only check if the angles get very
small. The fast method of computation of the angles is
developed in papers~\cite{FastHyp12, CLV2012}. Its
implementation for systems with a single time delay can be
found in Refs.~\cite{HypDelay2016}, and in
Ref.~\cite{HypManyDelay2018} the generalization for the case
of multiple delays is provided.

\begin{figure}
  \fighere{0.99}{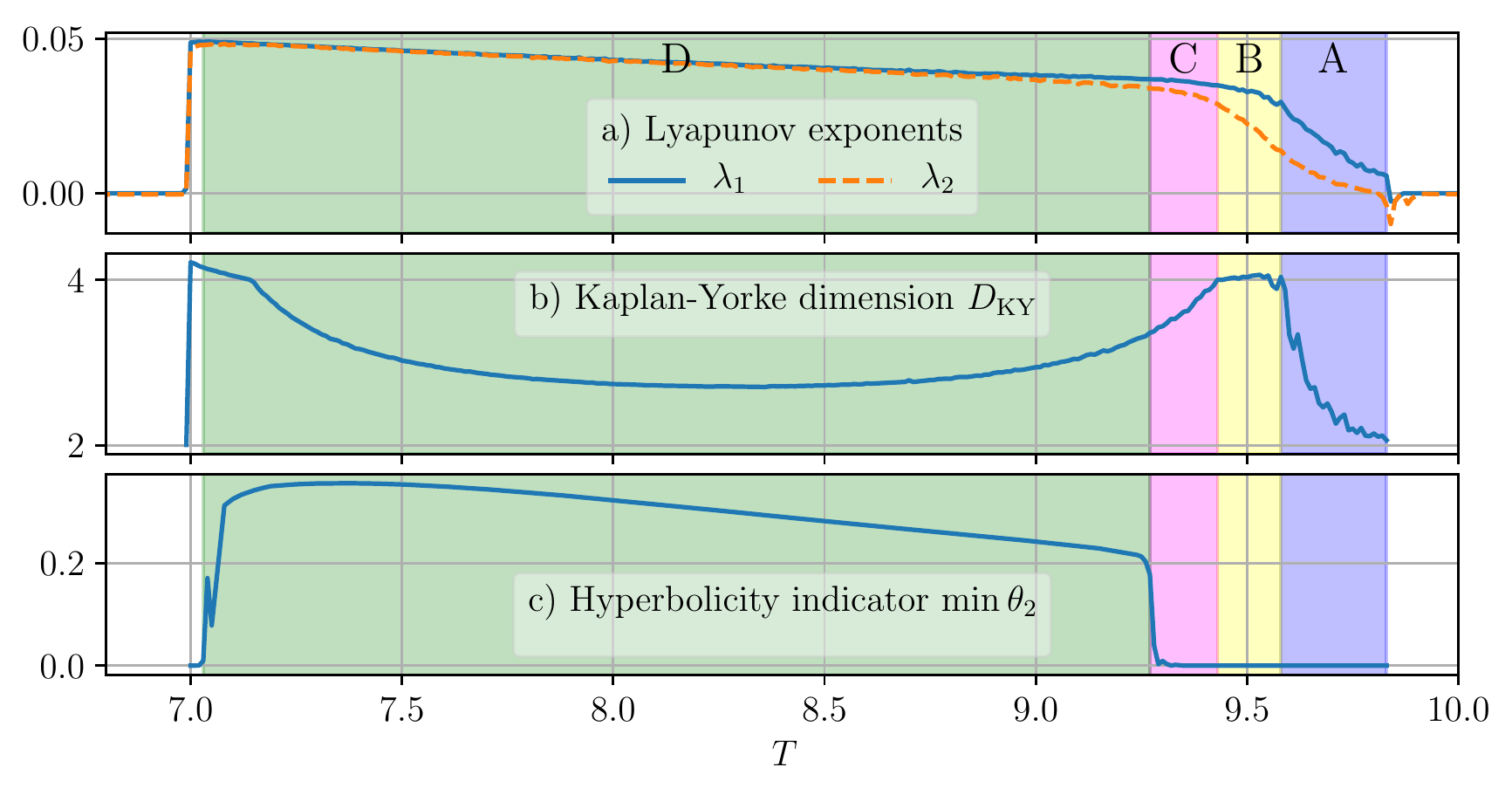}
  \caption{(a) Lyapunov exponents, (b)
    Kaplan-Yorke dimension and (c) Minimal angle $\theta_2$
    between expanding and contracting tangent subspaces.
    Shaded areas A, B, C and D highlight ranges of different
    attractor types.}
  \label{fig:anglam}
\end{figure}

A chaotic attractor is known to contain invariant subsets,
in particular, periodic orbits, and there are effective
numerical methods for detection these embedded
orbits~\cite{UPO99, UPO06, UPO09}. Nevertheless, application
of these methods for high-dimensional systems is rather
problematic yet. In this paper we develop another approach
to detect some of the embedded invariant subsets.

Running along a trajectory we can encounter points where
some of CLVs merge, i.e., the angle between them
vanishes. It can occur either for vectors from unstable and
stable subsets or for any other pair of vectors. Merging two
CLVs means that the number of CLVs in this point is less by
one compared to another attractor points. It means that we
have here some degenerated invariant subset. To provide a
simple illustration assume that the attractor contains a
fixed point with real eigenvalues. CLVs of this invariant
subset are merely its eigenvectors. Let two eigenvalues
coincide and the only one eigenvector corresponds to this
degenerated pair (for example this is always the case in two
dimensional space). Then this will be exactly the situation
described above. Thus running along an attractor trajectory
and collecting points where some of CLVs merge we will
detect degenerated invariant subsets embedded into the
attractor.

The degenerated subsets can be identified by a signature
constructed as a list of indexes of merging CLVs. Obviously
we can not distinguish two subsets with identical signatures
and they will be treated as a one subset. For each signature
we will compute CLV based FTLEs, i.e., average rates of
exponential growths or decay of CLVs near the subsets. These
partial FTLEs will be plotted against a control parameter to
demonstrate how the subsets are transformed.

\section{The analysis}
\label{sec:analysis}

\subsection{Lyapunov exponents, Kaplan-Yorke dimension and
  angles between tangent subspaces}
\label{subsec:lyap}

\begin{figure}
  \fighere{0.99}{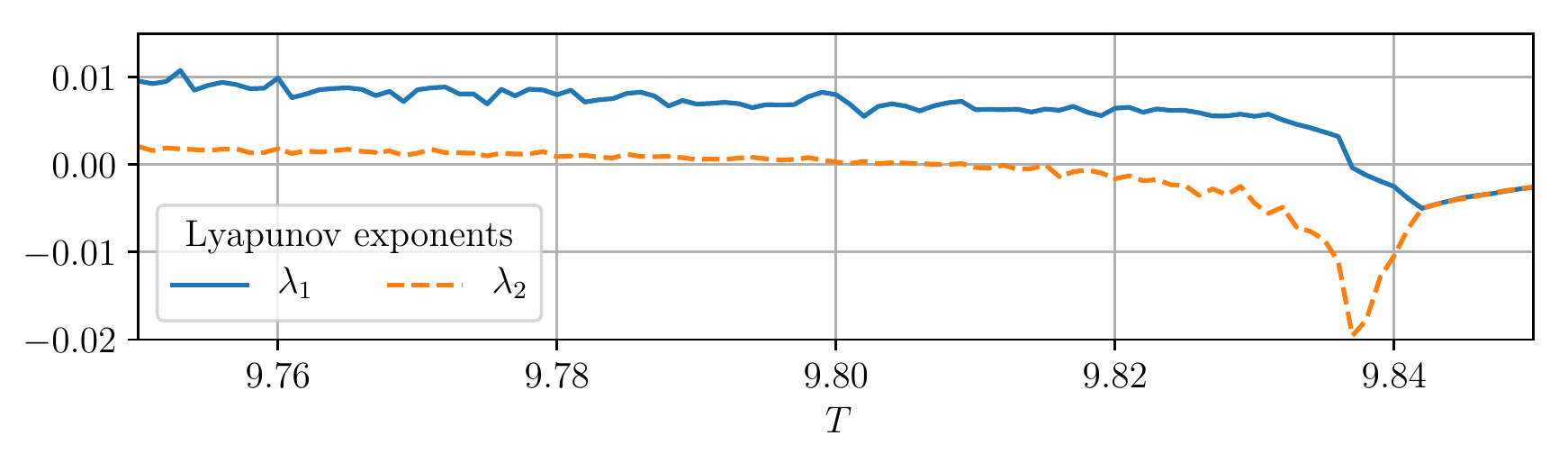}
  \caption{Enlarged area of
    Fig.~\ref{fig:anglam}(a) where the transition from chaos
    to hyperchaos occurs.}
  \label{fig:lammagn}
\end{figure}

First, for the system~\eqref{eq:sys} with
parameters~\eqref{eq:params} we consider minimal angle, two
Lyapunov exponents and Kaplan-Yorke dimension as functions
of $T$ decreasing from $10$ to $7$, see
Fig.~\ref{fig:anglam}. Near $T=10$ there are no positive
Lyapunov exponents, as one can see in
Fig.~\ref{fig:anglam}(a). It corresponds to regular
oscillations in the system.

Transition to chaos and then to hyperchaos is illustrated in
Fig.~\ref{fig:lammagn} where the enlarged area of
Fig.~\ref{fig:anglam}(a) is shown. One can see that the
dynamics becomes chaotic at approximately $T=9.835$. There
is a very narrow area where only one Lyapunov exponent is
positive, and very soon approximately at $T=9.8$ the second
one also becomes positive, so the hyperchaotic regime
appears.  Observe very small slope of the curve $\lambda_2$:
It goes almost horizontally when passes zero. This is
typical behavior for transition to hyperchaos owing to the
bifurcations of unstable periodic orbits embedded into
attractor~\cite{Yanchuk01}. In our case the area where
$\lambda_2$ stays close to zero is very narrow.

Parameter interval of our interest can be split into four
areas that we mark by capital letters A, B, C, and D, see
Fig.~\ref{fig:anglam}. Area A starts when the system becomes
hyperchaotic at $T=9.8$ and extends to $T=9.58$ until the
Kaplan-Yorke dimension grows,
see~Fig.~\ref{fig:anglam}(b). The minimal angle
$\min\theta_2$ between the expanding and contracting
subspaces remains close to zero (see
Fig.~\ref{fig:anglam}(c)) indicating non-hyperbolicity of
chaos within this area.  Area B covers the range where the
Kaplan-Yorke dimension is constant and ends at approximately
$T=9.43$. The minimal angle $\min\theta_2$ is still zero,
i.e. chaos is non-hyperbolic. Area C stretches up to a point
of transition to hyperbolic chaos, i.e., to a point where
the minimal angle $\theta_2$ starts to grow.  It occurs at
approximately $T=9.27$. Finally, area D corresponds to
hyperbolic hyperchaos with two positive Lyapunov
exponents. At the boundary of this area two first Lyapunov
exponents approach each other and merge,
see~Fig.~\ref{fig:anglam}(a). As discussed above the
considered system operates as two identical alternating
chaotic subsystems. The value of $T$ controls their relative
coupling strength with respect to the phase doubling
mechanism responsible for the hyperbolicity, see
Fig.~\ref{fig:operation}(b) and the related discussion. Thus
the hyperbolic hyperchaos in our system emerges when the
coupling strength between the subsystems becomes
sufficiently weak so that they demonstrate almost identical
dynamics. The area of hyperbolic hyperchaos ends at
approximately $T=7.02$. Beyond this point the system remains
hyperchaotic within the very narrow interval and then
oscillations become regular.

\subsection{Probability density functions on the
  attractor}
\label{subsec:pdf}

\begin{figure*}
  \figherewide{0.8}{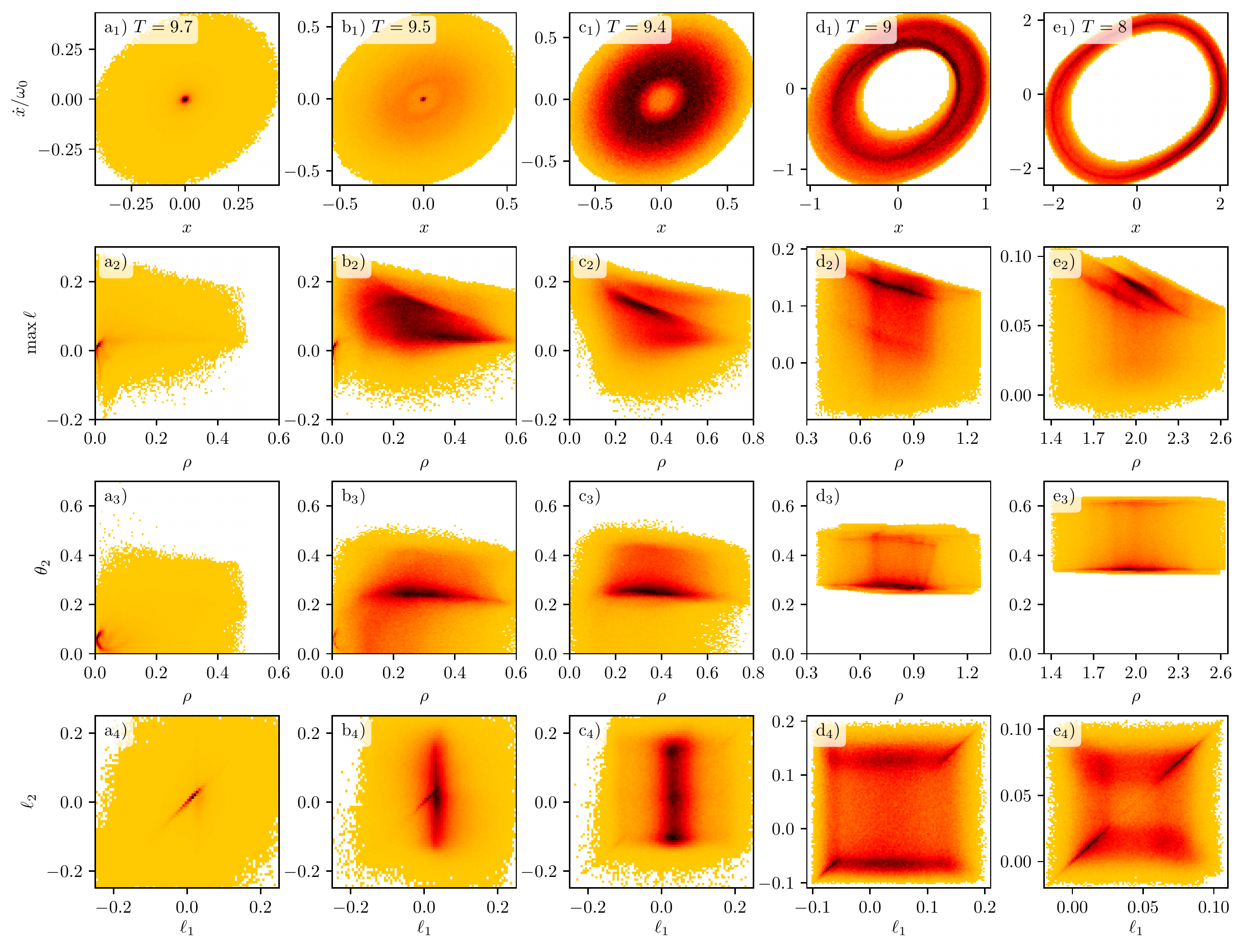}
  \caption{Numerical approximations of probability density
    functions (2D histograms) on the attractor. Columns (a),
    (b), and (c) correspond to areas A, B, and C in
    Fig.~\ref{fig:anglam}, respectively, and columns (d) and
    (e) correspond to the area D. Values of $T$ are given in
    the legends on the top row and are the same along
    the columns. Values on the vertical axis in the leftmost
    column are the same along rows. Darker areas represent
    higher densities.}
  \label{fig:attr}
\end{figure*}

To examine the attractor structure in areas A, B, C and D we
will consider now probability density functions (PDFs) of
dynamical variables and related characteristic values, see
Fig.~\ref{fig:attr}. Plots in this figure are grouped in
five columns. The first three of them, from (a) to (c),
correspond to the areas from A to C in
Fig.~\ref{fig:anglam}, and two last columns (d) and (e)
represent the area D. Column (d) characterizes a
hyperchaotic hyperbolic attractor close to the transition
point and (e) corresponds to the case when the
relation~\eqref{eq:hchaos_cond} is fulfilled exactly for
$k=2$, i.e., $\tau=1.5T$.  All plots are computed for the
stroboscopic map at $t=nT$.

\subsubsection{Area A}

Figures~\ref{fig:attr}(a$_{1,2,3,4}$) are plotted at $T=9.7$
that corresponds to the area A. Panel (a$_1$) shows PDF of
$x$ and $\dot x/\omega_0$. Since the phase space dimension
is high, these plots can be considered as two-dimensional
projections of multidimensional PDFs. Observe a sharp spike
at the origin visible as a dark spot. The spot is surrounded
by a wide pale area representing wandering of the system in
the vicinity of the origin.

\begin{figure}
  \fighere{0.99}{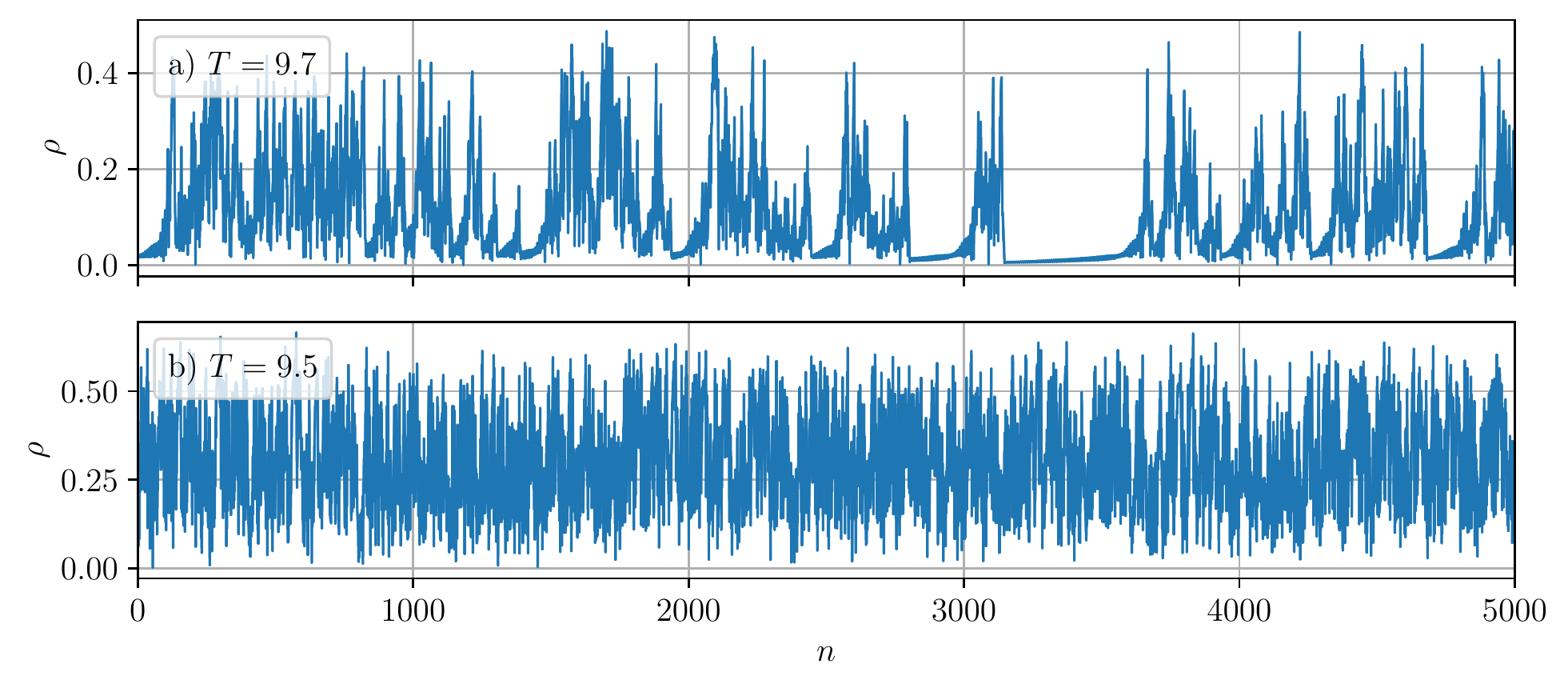}
  \caption{Time series of distances to the origin for (a)
    $T=9.7$, area A, (b) $T=9.5$, area B.}
  \label{fig:tser}  
\end{figure}

\begin{figure}
  \fighere{0.99}{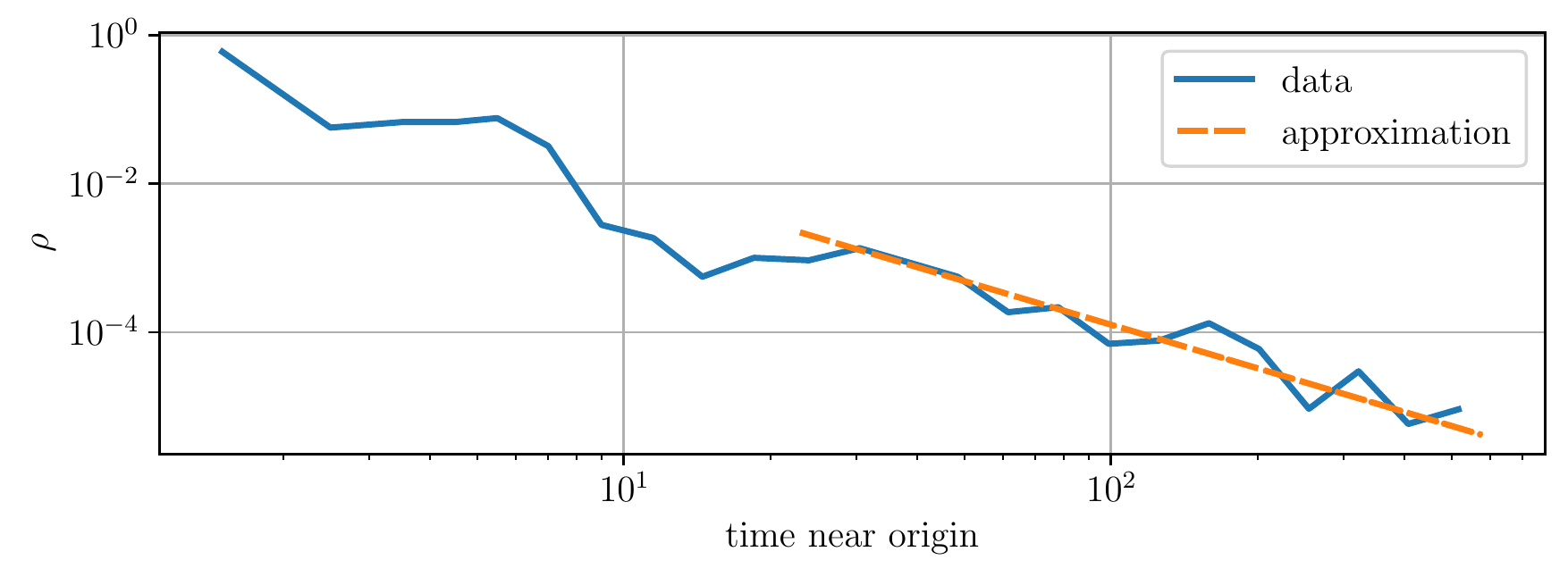}
  \caption{Distribution of laminar phases $\rho<0.05$ and
    its power law approximation with the exponent
    $\alpha=1.95$ that corresponds to
    Figs.~\ref{fig:attr}(a) and \ref{fig:tser}(a), area A.}
  \label{fig:intermpwlaw}
\end{figure}

The observed structure of PDF is caused by intermittency,
see Fig.~\ref{fig:tser}(a). In this figure we plot the phase
space distance of the orbit to the origin
\begin{equation}
  \label{eq:rho}
  \rho(t)=\sqrt{x(t)^2+(\dot x(t)/\omega_0)^2}
\end{equation}
for time sliced stroboscopically at $t=nT$. One can see
alternation of laminar phases when the system is close to
zero with burst of
oscillations. Figure~\ref{fig:intermpwlaw} provides further
confirmation of the intermittent nature of the considered
regime. It shows a distribution of lengths of laminar
trajectory cuts when the phase space trajectory is near the
origin. Here and below power law distributions as well as
estimation of the exponent $\alpha$ is done with the help of
Python package ``powerlaw''~\cite{powerlaw}.  In the log-log
scale the distribution admits linear approximation that
corresponds to the power law. The computed exponent is
$\alpha=-1.95$.

Figure~\ref{fig:attr}(a$_2$) shows PDF of $\rho$ and maximal
FTLE $\max \ell_i$, $i=1,2,\ldots$. Here and below FTLEs
$\ell_i$ are computed as average exponential growth rates of
CLVs over time $T$, corresponding to one step of the
stroboscopic map. Since FTLEs strongly fluctuate, they are
usually not ordered in the descent order in contrast to the
global Lyapunov exponents. Hence, on each step we simply
take the largest one.

The PDF of $\rho$ and $\max \ell$ in
Fig.~\ref{fig:attr}(a$_2$) shows locations of areas of
chaotic divergence on the attractor and areas where close
trajectories approach each other. One can see a spike at
$\rho=\max\ell=0$. It corresponds to the laminar phases and
indicates that near the origin the trajectories basically
demonstrate marginal stability. To explain this, we need to
take into account that $x=\dot x=0$ is a fixed point for the
considered system~\eqref{eq:sys}. Linearization near this
point results in a linear equation with parametric
excitation with the period $T$. Since the excitation
parameter oscillates symmetrically near zero, on average the
fixed point at the origin is marginally stable.

Figure~\ref{fig:attr}(a$_3$) shows PDF of $\rho$ and the
angle $\theta_2$ whose zero indicates a tangency between
expanding and contracting manifolds, i.e., reveals points of
the hyperbolicity violation. From Fig.~\ref{fig:anglam}(c)
we know that within the area A chaos is non-hyperbolic. And
from Fig.~\ref{fig:attr}(a$_3$) we see that the violation of
the hyperbolicity preferably occurs near the origin: one can
see the spike near $\rho=0$ where $\theta_2$ often
vanishes. Beyond the origin the vanishing angles are more
rare.

Figure~\ref{fig:attr}(a$_4$) represents PDF of two first
FTLEs $\ell_1$ and $\ell_2$. The dark line along the
diagonal near the origin corresponds to equal $\ell_1$ and
$\ell_2$. Identical Lyapunov exponents reflect a symmetry of
dynamics with respect to some variables interchange. As
discussed above, see Fig.~\ref{fig:operation}, within the
considered parameter range the system can be treated as two
weakly coupled chaotic subsystems. The stripe along the
diagonal in Fig.~\ref{fig:attr}(a$_4$) indicates that these
two subsystem behave coherently, i.e., synchronized, when
pass the origin.

\subsubsection{Area B}

PDF of $x$ and $\dot x/\omega_0$ in the area B, see
Fig.~\ref{fig:attr}(b$_1$), looks very similar to the
previous case in Fig.~\ref{fig:attr}(a$_1$). The only
difference is a barely visible darker area surrounding the
origin. But actually it indicates a qualitative change of
the behavior. One can see in Fig.~\ref{fig:tser}(b) that
though a trajectory still often visits the origin
neighborhoods, this is not an intermittency.

PDF of $\rho$ and $\max \ell$ in Fig.~\ref{fig:attr}(b$_2$)
reveals the emergence in the area B of a new structure. Like
in the area A, see Fig.~\ref{fig:attr}(a$_2$), we observe
the spike at $\rho=\max\ell=0$ corresponding to the passing
of a trajectory near the fixed point at the origin. But also
a massive bulk of points appears at positive $\max \ell$ and
nonzero $\rho$. It represents a chaotic subset embedded into
the attractor. Thus in the area B the dynamics is determined
by a wandering of the system between this chaotic subset and
the fixed point at the origin.

PDF of $\rho$ and $\theta_2$ in Fig.~\ref{fig:attr}(b$_3$)
again, similarly to the area A, see
Fig.~\ref{fig:attr}(a$_3$,), contains the spike near
$\rho=\theta_2=0$ (now barely visible due to the presence of
another maxima) but also a large spot corresponding to a new
chaotic subset. Most of points within this spot are
hyperbolic, i.e., located at $\theta_2>0.2$. However, their
noticeable number is characterized by a vanishing angle:
observe getting down to $\theta_2=0$ darker arm centered at
approximately $\rho=0.2$.

PDF of $\ell_1$ and $\ell_2$, see
Fig.~\ref{fig:attr}(b$_4$), again contains the diagonal line
near zero mentioned already in the area A in
Fig.~\ref{fig:attr}(a$_4$) and corresponding to the
coherence of the subsystems near the origin. Also a very
well pronounced is the vertical stripe representing strong
fluctuation of the second FTLE $\ell_2$. The first one
$\ell_1$, on contrary, is well localized.

Altogether, in the area B a new embedded non-hyperbolic
chaotic subset emerges but the fixed point at the origin is
essential yet and trajectories wander between these two
subsets. Due to this wandering, the FTLEs $\ell_1$ and
$\ell_2$ switch between coherency at the origin and strong
fluctuation of $\ell_2$ at the chaotic subset. As shown
below, the presence of two competing embedded subsets
results in anomalous diffusion of Lyapunov exponents.

\subsubsection{Area C}

Figure~\ref{fig:attr}(c$_1$) corresponds to the area C. One
can see that darker and barely visible circular structure in
Fig.~\ref{fig:attr}(b$_1$) is now turned into a well formed
ring where trajectories spend most of time. There is no
maximum corresponding to the fixed point at the origin
anymore, but its neighborhood is still visited.

In Fig.~\ref{fig:attr}(c$_2$) we observe that the spike at
$\rho=\max\ell=0$ disappears at all, and the origin $\rho=0$
is characterized by a positive FTLE. Thus, the chaotic subset
first appeared in Fig.~\ref{fig:attr}(b$_2$) now
dominates. Representing it structure on the PDF becomes
sharper in comparison with Fig.~\ref{fig:attr}(b$_2$): one
can see well defined dark stripe on the plot indicating that
the most probable largest FTLEs decrease as $\rho$ grows.

Figure~\ref{fig:attr}(c$_3$) also confirms disappearance of
the structure representing the fixed point at the
origin. The dominating chaotic subset becomes ``more
hyperbolic'' insofar as the getting down arm near $\rho=0.2$
disappears and less number of points has the vanishing angle
$\theta_2$.

In accordance with the changed role of the fixed point, in
Fig.~\ref{fig:attr}(c$_4$) no diagonal stripe is visible
representing the coherence of $\ell_1$ and $\ell_2$ at the
origin. The only most visited structure is a vertical stripe
corresponding to the chaotic subset that now
dominates. Moreover barely visible are two more
features. The first is a pair of diagonal segments appeared
beyond the origin at approximately
$\ell_1=\ell_2\approx \pm 0.15$, and the second are darker
areas to the left and to the right from the main vertical
stripe representing a more intense fluctuations of
$\ell_1$. These features are precursors of a hyperchaotic
hyperbolic attractor that appears in the course of further
decrease of $T$.

\subsubsection{Area D}

Figures~\ref{fig:attr}(d) and (e) represent a hyperchaotic
hyperbolic attractor, the area D in
Fig.~\ref{fig:anglam}. The common feature is that
trajectories never visit vicinities of the origin, see
Figs.~\ref{fig:attr}(d$_1$ and e$_1$). It means that
oscillation phase that is responsible for the hyperbolic
chaos is now well defined~\cite{BaranovHyper10}.

Inspection of PDFs in Figs.~\ref{fig:attr}(d$_1$ and e$_1$)
reveals two different forms of the attractor. At $T=9$,
i.e., just after the transition to the hyperbolicity, there
are two loops formed by the most visited points while far
from the transition at $T=8$, only one main loop is
visible. We recall that hyperchaos in the considered
system~\eqref{eq:sys} is the result of interaction of two
coupled chaotic subsystems, see Fig.~\ref{fig:operation}(b)
and the related discussion. When $T$ is decreased the
coupling strength between the subsystems becomes weaker
while the hyperbolicity mechanism, related with the phase
doubling, becomes stronger. Comparing
Figs.~\ref{fig:attr}(d$_1$) and (e$_1$) we can see that it
results in more coherent behavior of these subsystems that
manifests itself as a merge of the two loops.

Visually the attractor in Fig.~\ref{fig:attr}(d$_1$) looks
more complicated then the attractor in
Fig.~\ref{fig:attr}(e$_1$) and one can expect that its
dimension is higher. This intuition agrees with
Fig.~\ref{fig:attr}(b): the Kaplan-Yorke dimension decreases
as $T$ is varied within the area D from $T=9$ to $T=8$.

PDF of $\rho$ and $\max\ell$ has a single stripe when the
coupling between the chaotic subsystems is stronger at
$T=9$, see Fig.~\ref{fig:attr}(d$_2$) while weaker coupling
at $T=8$ results in two parallel stripes. The latter
indicates that two chaotic subsets may be distinguished, in
the phase space and a trajectory wanders between
them. Taking into account the discussion of
Figs.~\ref{fig:attr}(d$_1$ and e$_1$) that at $T=8$ the two
subsystems are more coherent, one can assume that these
subsets correspond to synchronized and non-synchronized
segments of trajectories. Unlike the case in
Fig.~\ref{fig:attr}(b$_2$) these two subsets have similar
properties so that, as we will discuss below, their presence
does not affect the normal convergence of Lyapunov
exponents. Also notice that most of points in
Figs.~\ref{fig:attr}(d$_2$ and e$_2$) are located at
positive values of $\max\ell$. It means that already locally
the fluctuations of FTLEs are sufficiently weak.

Also the emergence of the two subsets at $T=8$ is visible in
PDFs of $\rho$ and $\theta_2$, see
Figs.~\ref{fig:attr}(d$_3$ and e$_3$). In the panel d$_3$ we
observe the main dominating structure as a dark horizontal
stripe at the bottom of the plot, while in the panel (e$_3$)
one also can see one more horizontal stripe at the top
part. It means that the two chaotic subsets has different
distributions of angles between expanding and contracting
manifolds though in both cases the angles are
non-vanishing. The latter again confirms the hyperbolicity
of chaos in the area D.

In PDF for $\ell_1$ and $\ell_2$ in
Fig.~\ref{fig:attr}(d$_4$) the features barely visible in
Fig.~\ref{fig:attr}(c$_4$) are well pronounced. Both of
FTLEs fluctuate with almost identical amplitudes. Both
positive and negative values are encountered but the
fluctuations are strongly biased to the positive
side. Prevailing structures are two horizontal stripes
representing fluctuations of $\ell_1$ and two short diagonal
segments corresponding to synchronous oscillations of the
two subsystems. Since the presence of the synchronized
segments is revealed only in PDF of $\ell_1$ and $\ell_2$,
we conclude that at $T=9$ the coherency between the
subsystems occurs but it is week and seldom.

As $T$ is decreased up to $8$, the PDFs of $\ell_1$ and
$\ell_2$ become symmetric with respect to the main diagonal,
see Fig.~\ref{fig:attr}(e$_4$). It demonstrates presence of
two identical chaotic subsystems that can oscillate
coherently, see the well formed diagonal stripe. But they do
not stay synchronized for all time. Ends of the coherent
stages associated with leaving the symmetric attractor are
represented by symmetric off-diagonal structures. Also
notice that the whole area of FTLEs fluctuations is roughly
four times narrower than in the previous cases and
fluctuating FTLEs are preferably positive.

\subsection{Degenerated invariant subsets}
\label{subsec:subsets}

\begin{figure*}
  \figherewide{0.8}{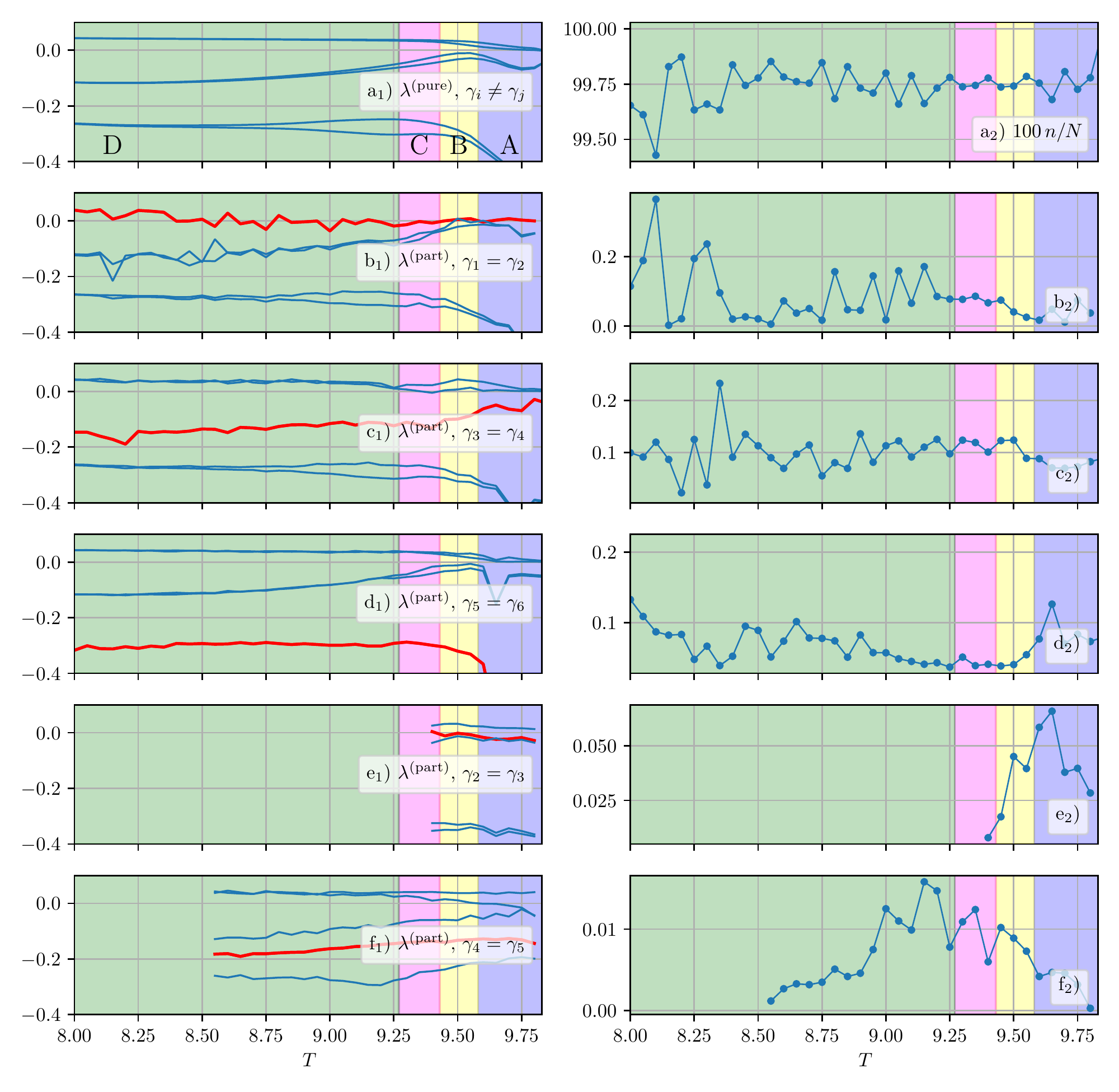}
  \caption{\label{fig:embed_lam}Left panel (a$_1$): Lyapunov
    exponents computed for trajectory points where no CLV
    coincide. Other panels below it: partial Lyapunov
    exponents computed for degenerated invariant subsets
    where some of CLVs coincide, see panel legends. Fat red
    lines highlight the identical partial Lyapunov exponents
    corresponding to the merged CLVs.  Right panel (a$_2$)
    and the panels below: relative size $100 n/N$ of the
    subset for that the corresponding left panel is
    plotted. Here $N=10^6$ is the total trajectory length,
    and $n$ is the number of encountered subset points.  }
\end{figure*}

When we move along an attractor trajectory and compute CLVs
$\gamma_i$ we can register cases when some of the vectors
merge. This happens when we pass close to a degenerated
invariant subset with smaller number of independent CLVs.
Such subsets can be identified by signatures of the form
$\gamma_i=\gamma_j$ indicating what vectors coincide, and
characterized by partial Lyapunov exponents
$\lambda^{\text{(part)}}$ computed as averages of the
corresponding FTLEs near these subsets.

\subsubsection{Partial Lyapunov exponents for the
  degenerated subsets}

Figure~\ref{fig:embed_lam} represents partial Lyapunov
exponents (left column) and their percentage (right column)
for the degenerated subsets. The latter refers to a number
$n$ of encountered points with a certain signature divided
by the total number of the checked attractor points $N=10^6$
and multiplied by $100$ . We show signatures with noticeable
percentages only. Those with $n\leq 50$ are omitted.

Figures~\ref{fig:embed_lam}(a) represent trajectory points
without peculiarities, i.e., those where no merging of CLVs
occurs. As one can see in panel (a$_2$) their percentage is
close to $100$ so that the curves in
Fig.~\ref{fig:embed_lam}(a$_1$) almost coincide with the
ordinary global Lyapunov exponents,
cf. Fig.~\ref{fig:anglam}(a).

Figures~\ref{fig:embed_lam}(b), (c) and (d) demonstrate the
largest degenerated subsets. Their signatures are
$\gamma_1=\gamma_2$, $\gamma_3=\gamma_4$, and
$\gamma_5=\gamma_6$, respectively. Their percentages are
around $0.1\div 0.2$, see panels (b$_2$), (c$_2$), and
(d$_2$). Observe that partial Lyapunov exponents in the
panels (b$_1$), (c$_1$), and (d$_1$) are similar to those
without merging CLVs in Fig.~\ref{fig:embed_lam}(a$_1$). The
common feature of these three cases is that the exponents
are either pairwise coincide or close to each other: 1 and
2, 3 and 4, 5 and 6. This is the manifestation of the
presence of two chaotic subsystems, see
Fig.~\ref{fig:operation}(b) and the related
discussion. Pairwise closeness of the exponents is related
with the coherence of these subsystems, discussed above.

The percentage of the subset $\gamma_2=\gamma_3$ in
Figs~\ref{fig:embed_lam}(e) is smaller in order of
magnitude. However, this subset is nevertheless
essential. We recall that the whole attractor has
two-dimensional expanding manifold so that the vanishing
angle between the second and the third CLVs indicates the
destruction of the hyperbolicity. Thus
Fig.~\ref{fig:embed_lam}(e$_1$) represents the subset
responsible for the violation of the hyperbolicity. Observe
that it disappears in the middle of the area C, before the
expected transition to hyperbolicity on the C-D
boundary. This is explained in
Fig.~\ref{fig:embed_lam}(e$_2$). One can see that the number
of the encountered points monotonically decays within the
area C as $T$ approaches the area D. The system visits the
subset $\gamma_2=\gamma_3$ more and more seldom so that one
have to trace longer and longer trajectories to detect this
subset close to the transition point.

Also notice that the destructing hyperbolicity subset
$\gamma_2=\gamma_3$ has the largest percentage in the area
A. This agrees well with the previous observations in
Figs.~\ref{fig:attr}(a$_3$ - e$_3$): The highest maximum of
PDF near $\theta_2=0$ indicating the disappearance of the
hyperbolicity is observed in the area of intermittency A.

The percentage of the subset $\gamma_4=\gamma_5$ in
Figs~\ref{fig:embed_lam}(f) is also relatively small. This
subset becomes hyperchaotic in the area B, and two largest
Lyapunov exponents become almost identical in the area D,
when the whole system becomes hyperbolic. Since its second
and third CLVs do not merge, this subset is hyperbolic. Also
it is located far from the symmetric manifold where the two
subsystems are synchronized since their partial Lyapunov
exponents are not close pairwise as in
Figs.~\ref{fig:embed_lam}(b, c, and d). It explains the
existence of two forms of hyperbolic attractors that where
demonstrated in Fig.~\ref{fig:attr}(d) and (e): Less
coherent attractor (column (d)) exists to the right of
$T\approx 8.5$ when the subset in
Figs~\ref{fig:embed_lam}(f) is visited, and it becomes more
coherent (column (e)) when this subset disappears.

\subsubsection{Bias of global Lyapunov exponents due to the
  degenerated subsets}

\begin{figure}
  \fighere{0.99}{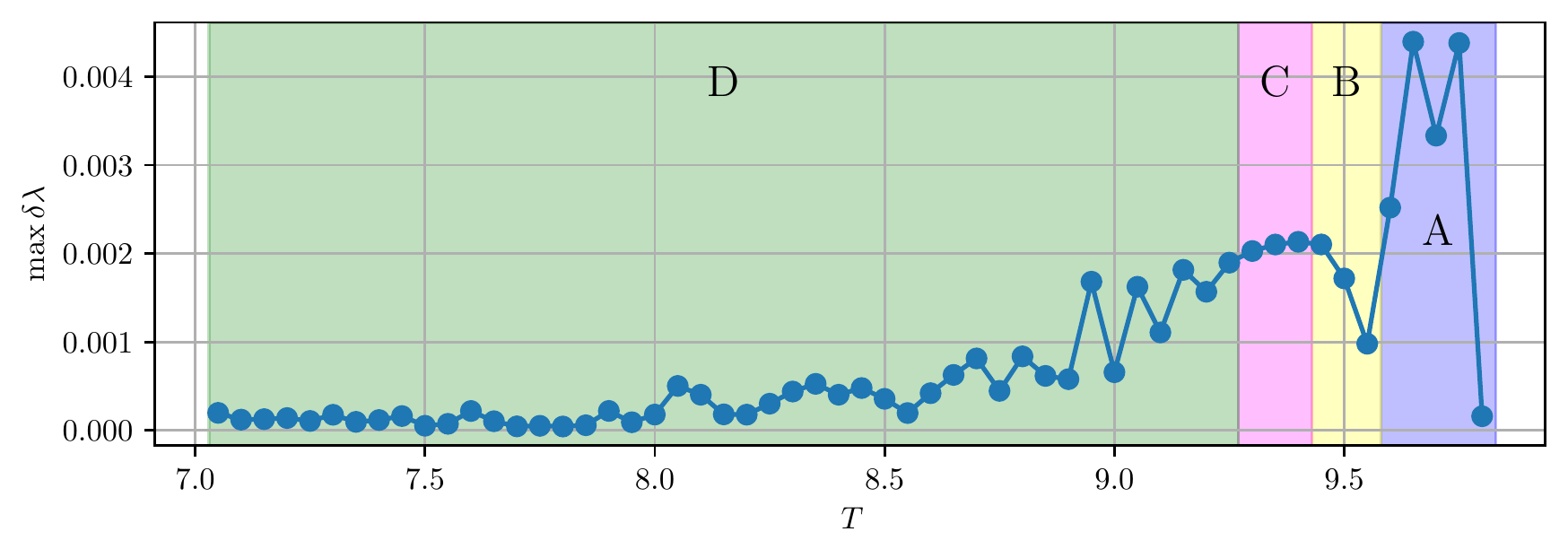}
  \caption{\label{fig:embed_pert}Maximal relative bias of
    Lyapunov exponents due to the degenerated subsets
    $\max\{\delta \lambda_i|i=1,2,\ldots,6\}$, see
    Eq.~\eqref{eq:lyap_bias}.}
\end{figure}

Now we estimate overall influence of the degenerated
subsets. For each $T$ we first compute the Lyapunov
exponents $\lambda_i$ as average of all CLV FTLE encountered
along a trajectory. Then we compute purified Lyapunov
exponents $\lambda^{\text{(pure)}}_i$ ignoring those FTLEs
obtained at points near the degenerated subsets. Finally we
estimate the relative bias introduced by the subsets as
\begin{equation}
  \label{eq:lyap_bias}
  \delta\lambda_i = \left|\frac{\lambda_i -
      \lambda^{\text{(pure)}}_i}
    {\lambda_i}\right|.
\end{equation}

Figure~\ref{fig:embed_pert} shows the maximal relative bias
computed for six Lyapunov exponents as
$\max\{\delta \lambda_i|i=1,2,\ldots,6\}$. Observe clear
difference of areas A, B, C and D. In the area A, where the
system demonstrates intermittency, the bias is the
highest. The area B is characterized by the presence of two
competing subsets embedded into the attractor. In the
beginning of this area, the bias first drops down but then
grows again.  In the area C, where the system has chaotic
non hyperbolic attractor, the bias remains at a constant
level.

In the area D, where the attractor becomes hyperbolic, we
observe a decrease of the bias. It occurs until the subset
$\gamma_4=\gamma_5$ is visited, see
Figs~\ref{fig:embed_lam}(f), so that the hyperbolic
attractor has the less coherent form shown in
Figs.~\ref{fig:attr}(d). After the disappearance of this
subset, when the attractor becomes more coherent (see
Figs.~\ref{fig:attr}(e)) the bias stays at more or less
constant small level. Small bias due to the degenerated
subsets is related with structural stability and uniformity
of the hyperbolic attractor, see
Refs.~\cite{KuzUFN11,HyperBook12} for the discussion of
these concepts.

\subsection{Large time FTLEs}
\label{subsec:diffus}

\begin{figure}
  \fighere{0.99}{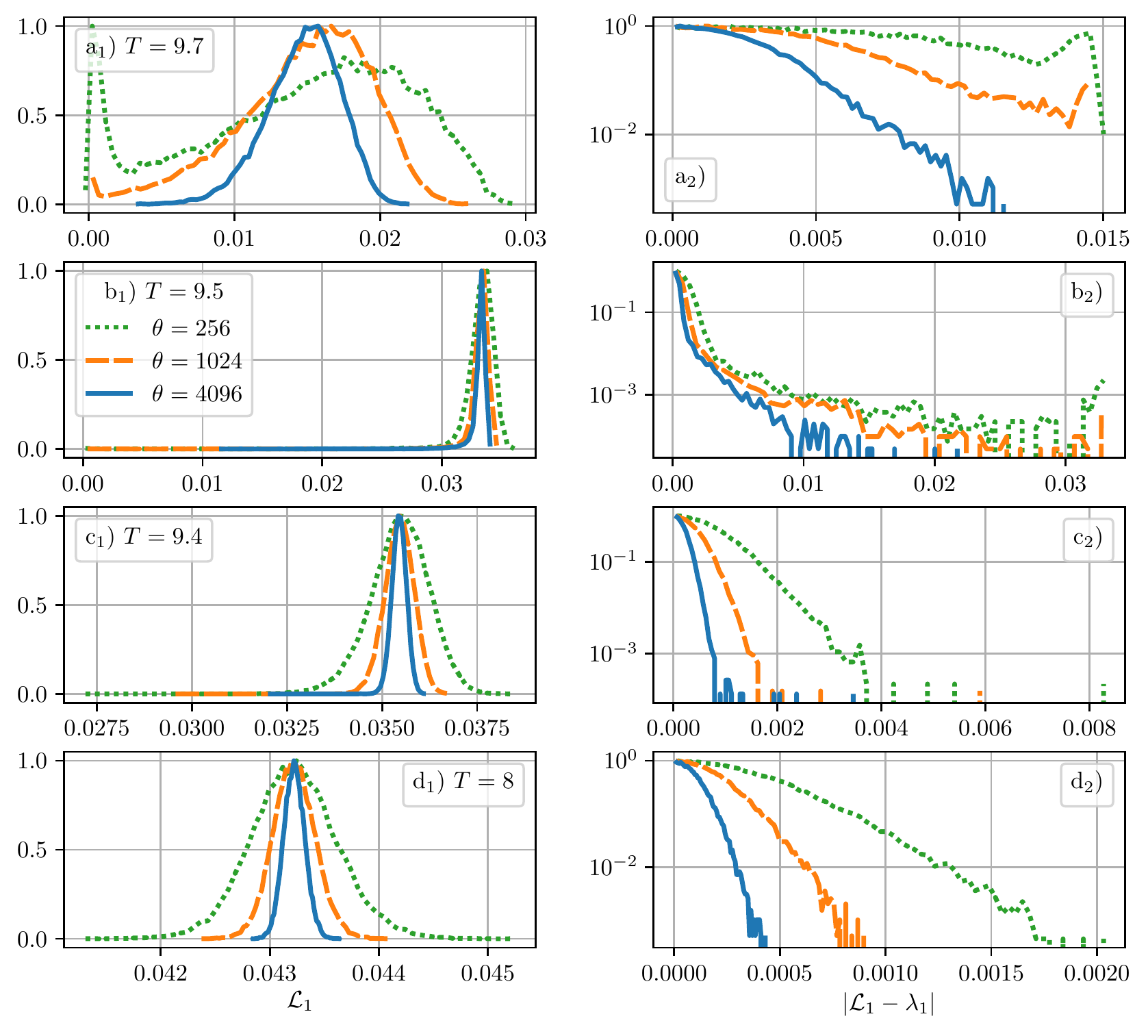}
  \caption{\label{fig:lamdistr} PDFs of large time FTLE
    $\mathcal{L}_1$ computed on increasing time scales
    $\theta=256$, $1024$, and $4096$. Log scale is used for
    the vertical axes in the right column.}
\end{figure}

Another approach considers fluctuations of FTLEs on large
times. Large time FTLEs averaged over $\Delta t=\theta T$
time intervals will be denoted as $\mathcal{L}_i$.

When oscillations are chaotic or hyperchaotic, computing
large time FTLEs we deal with sums of nearly independent
random values~\cite{Fujisaka83, StatMechLyap11}. It means
that PDF of large time FTLEs is expected to be Gaussian, and
the summation can be treated as a diffusion process with
linear growth of dispersion~\cite{Fujisaka83,
  StatMechLyap11}. When this is indeed the case, the
diffusion of Lyapunov exponents is said to be
normal. Otherwise it is anomalous. Below we will see that
the considered system can demonstrate behaviors of both
types.

\subsubsection{Distributions of large time FTLE}

Figure~\ref{fig:lamdistr} shows PDFs for the first large
time FTLE computed at $\theta=256$, $1024$, and $4096$. The
left column represents $\mathcal{L}_1$ itself and the right
one is for absolute values of deviations of $\mathcal{L}_1$
from the mean $\langle\mathcal{L}_1\rangle=\lambda_1$. In
the right column logarithmic scale is used on the vertical
axis. 

Figure~\ref{fig:lamdistr}(a) corresponds to the area A,
where the system demonstrates intermittency and laminar
phases obeys power law, see Fig.~\ref{fig:intermpwlaw}.
Hence laminar phases of arbitrary lengths can appear with a
nonzero probability, and small $\mathcal{L}_1$ can be
encountered regardless of $\theta$. As a result the
corresponding PDF of $\mathcal{L}_1$ always has a nonzero
but asymptotically decaying peak at the origin. This is
illustrated in Fig.~\ref{fig:lamdistr}(a$_1$). One can see
that this peak is very high at $\theta=256$, and it is still
visible at $\theta=1024$. The left tail of the PDF at
$\theta=4096$ does not reach the origin, but this is because
the number of points accumulated for computation of PDF is
not enough to take into account very long laminar phases.

If we ignore the left end of the curve, the rest looks
Gaussian. The main its feature, the exponentially decaying
tails, is confirmed in Fig.~\ref{fig:lamdistr}(a$_2$). One
can see here that in the logarithmic scale on the vertical
axis the tails of PDFs at $\theta=1024$ and $\theta=4096$
decay linearly that corresponds to the exponential law. This
Gaussian form correlates with the uniform distribution of
the one step FTLE $\ell_1$ outside of the origin, see
Fig.~\ref{fig:attr}(a$_4$).

\begin{figure}
  \fighere{0.99}{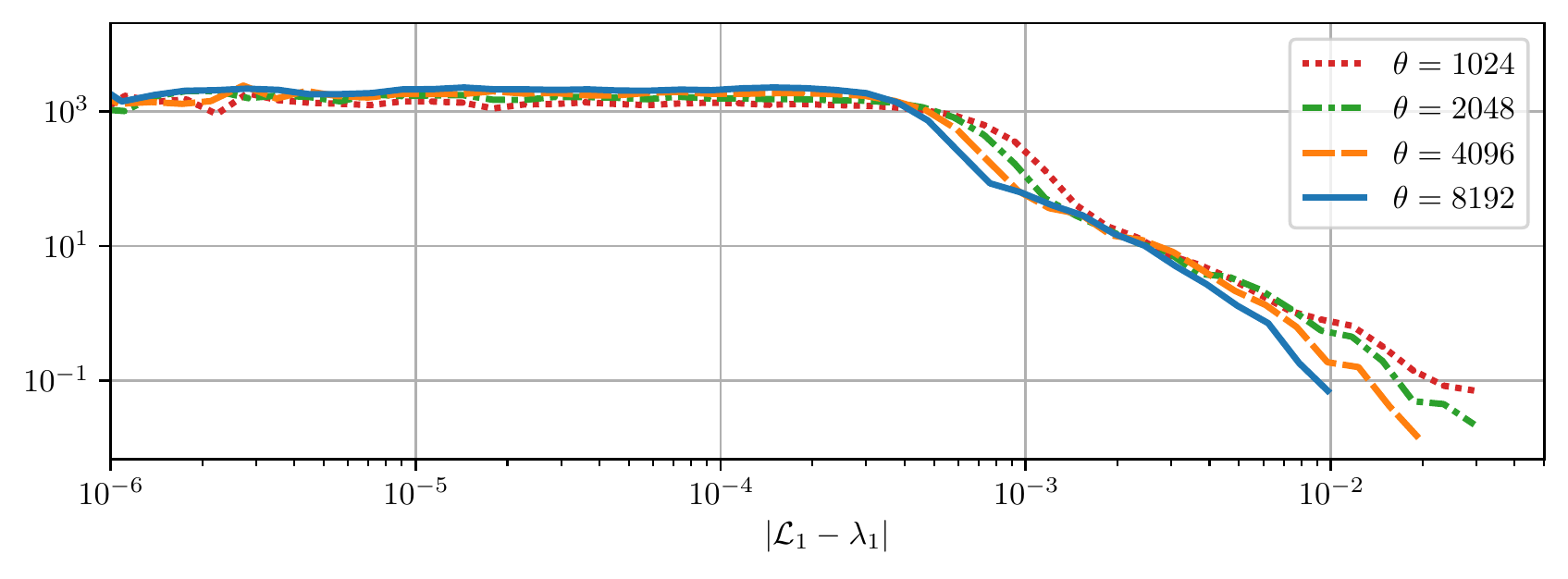}
  \caption{\label{fig:pwtails} PDF for deviations of
    $\mathcal{L}_1$
    on increasing time
    scales for $T=9.5$. Log-log scale is used. Observe power-law decay
    of tails. 
    Estimated values of the exponent $\alpha$ are $(\theta, \alpha)$ =
    (1024, 2.24), (2048, 2.28), (4096, 3.16), (8192, 2.72)}
\end{figure}

In the area B there are two competing structures in the
phase space: one is the fixed point at the origin and the
other is a chaotic subset, see
Fig.~\ref{fig:attr}(b$_2$). Their stability properties are
strongly different. Wandering between them results in non
Gaussian distributions of FTLEs. One can see in
Fig.~\ref{fig:lamdistr}(b$_1$) that regardless of the
averaging time there is an essential tail spreading to the
origin. This tail is also shown in
Fig.~\ref{fig:lamdistr}(b$_2$). One can see here that the
deviations decay is essentially slower then the exponent. To
further characterize PDF in this case we re-plot it in
Fig.~\ref{fig:lamdistr}(b$_2$) in log-log scale, see
Fig.~\ref{fig:pwtails}. Linear decay of the tails indicate
power law distribution, known as distribution with heavy
tails. The exponents of this distribution $\alpha$ are given
in the figure caption. One can see that it approaches to $3$
as $\theta$ grows.

The PDFs in the areas C (non hyperbolic chaos) and D
(hyperbolic chaos) demonstrate plain behaviors, typical for
common chaotic dynamics. The curves are Gaussian, see
Figs.~\ref{fig:lamdistr}(c$_1$) and (d$_1$) and their tails
decay exponentially, see Figs.~\ref{fig:lamdistr}(c$_2$ and
d$_2$). Notice that this is observed regardless of the
presence of the hyperbolicity: The Gaussian curves are
formed because the attractor does not contain strongly
competing subsets with different stability properties, and
the subset responsible for the violation of the
hyperbolicity has small relative weight,
see. Fig.~\ref{fig:embed_lam}(e).

\subsubsection{Diffusion of Lyapunov exponents}

\begin{figure}
  \fighere{0.99}{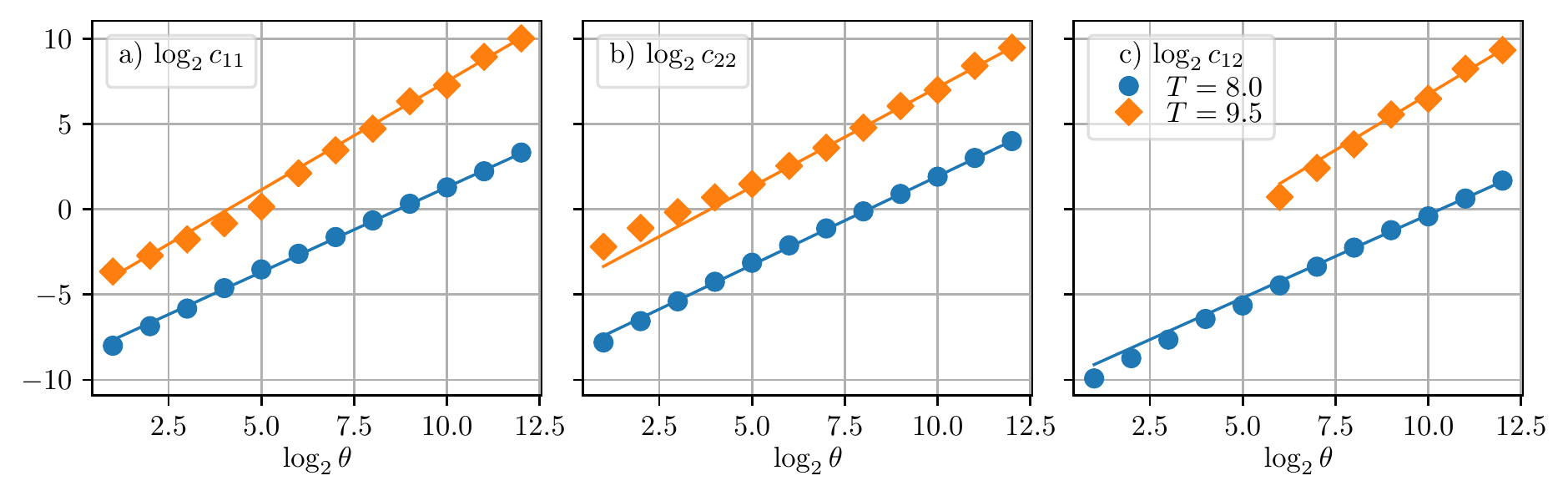}
  \caption{\label{fig:pwlawapprox} Power law approximation $D
    \theta^\sigma$, see Eqs.~\eqref{eq:pwcovar_approx_80} and
    \eqref{eq:pwcovar_approx_95} for numerical values. Only four last
    points are used for computing the approximations.}
\end{figure}

Summation of FTLEs in the course of computation of Lyapunov
exponents can be considered as a sort of random walking. If
the summed values are random and independent, the variance
of the sum is known to grow linearly in time. The
coefficient of this growth, a diffusion coefficient, can be
considered as one more characteristic quantifier of chaotic
dynamics~\cite{Fujisaka83, StatMechLyap11}. This is the case
for the most of chaotic systems no matter hyperbolic or not,
and it is related with the Gaussian form of PDFs of FTLEs on
large times, see Figs.~\ref{fig:lamdistr}(c) and (d). But
for non Gaussian distributions, like the one shown in
Fig.~\ref{fig:lamdistr}(b), an anomalous diffusion occurs.

Let $L_i(\theta)=\mathcal{L}_iT\theta$ be a Lyapunov sum
over $\theta$ steps, and let
$c_{ij}=\Covar[L_i(\theta), L_j(\theta)]$ be a covariance of
two Lyapunov sums. To reveal anomalous diffusion we
approximate the time dependence of the covariance via a
power law
\begin{equation}
  \label{eq:pwcovar}
  c_{ij}=D\theta^\sigma.
\end{equation}
For a normal diffusion $\sigma=1$ and $D$ is a usual
diffusion coefficient. In Fig.~\ref{fig:pwlawapprox} markers
show $c_{ij}$ for $i,j=1,2$ as functions of $\theta$ in
log-log scale. Solid lines represent computed
approximations. Line for $c_{12}$ at $T=9.5$ in
Fig.~\ref{fig:pwlawapprox}(c) starts beyond the origin since
$c_{12}$ for smaller $\theta$ are negative. Observe that
points are very good fitted to straight lines in the log-log
scale. Computed approximation at $T=8$ are
\begin{equation}
  \label{eq:pwcovar_approx_80}
  c_{11}=0.0025\,\theta^{1.00},
  c_{22}=0.0028\,\theta^{1.04},
  c_{12}=0.00093\,\theta^{0.98},
\end{equation}
and at $T=9.5$ are
\begin{equation}
  \label{eq:pwcovar_approx_95}
  c_{11}=0.027\,\theta^{1.27},
  c_{22}=0.044\,\theta^{1.17},
  c_{12}=0.012\,\theta^{1.31}.
\end{equation}

We see that in the area D of hyperbolic chaos at $T=8$ the
diffusion is normal: $\sigma\approx 1$. The diffusion
coefficients $D$ are of the order $10^{-3}$. For a non
Gaussian case $T=9.5$ in the area B $\sigma$ is larger then
$1$. It means that here we have anomalous diffusion. But
since $\sigma$ is nevertheless close to $1$, the comparison
of $D$ with the case at $T=8$ makes sense yet. We see that
it is of order $10^{-2}$, i.e., one order higher. Altogether
it indicates much higher amplitude of fluctuations of FTLEs
in the non Gaussian case.

\section{Outline and discussion}
\label{sec:outline}

In this paper we have considered a non-autonomous time delay
system whose excitation parameter is periodically modulated
so that the system produces a sequence of oscillation
pulses. Due to specially tuned nonlinear mechanism, phase of
the oscillations is doubled after each modulation period. As
a result a stroboscopic map for this system demonstrates
hyperbolic chaos. Varying relation between the delay time
and the excitation period one can observe a transition to
regime a when this map operates as two weakly coupled
chaotic subsystems excited alternately. The overall dynamics
in this case still being hyperbolic becomes hyperchaotic
with two positive Lyapunov exponents.

We have analyzed the transition to this hyperbolic
hyperchaos and reveled the following scenario. After regular
oscillations the hyperchaos appears almost immediately. An
area with a single positive exponent is very narrow. Then,
the following hyperchaotic regimes take place sequentially:
(a) intermittency as an alternation of staying near a fixed
point and chaotic bursts; (b) competition between the fixed
point and chaotic subset which appears near it; (c) plain
hyperchaos without hyperbolicity after termination visiting
neighborhoods of the fixed point; (d) transformation of
chaos to hyperbolic form.

The competition in the regime (b) results in a non-Gaussian
distribution of large time FTLE with power law tails and
power law growth of Lyapunov sums. This type of behavior
related with wandering of trajectories near subsets with
different numbers of expanding directions is called unstable
dimension variability (UDV). Usually it is observed as a
part of scenario of destruction of chaotic synchronization
of two subsystems~\cite{KapMaist00}. In our case we also can
talk about two chaotic subsystems with rather non-trivial
interaction. The UDV effect is observed for them as their
effective coupling strength is decreased.

The transition to hyperbolic hyperchaos (d) is accompanied
by vanishing of the embedded into the attractor
non-hyperbolic chaotic subset, that we have detected using
covariant Lyapunov vectors. The hyperbolic hyperchaos in
turn is found to be of two types. The difference is due to
the presence of the degenerated hyperbolic chaotic
subset. When it is visited by trajectories, the attractor
gets more complicated structure with higher Kaplan-Yorke
dimension, and after its vanish the system operates just as
two weakly coupled identical hyperbolic chaotic subsystems.

\begin{acknowledgments}
  Work of PVK on theoretical formulation, elaboration of
  computer routines and numerical computations was supported
  by grant of Russian Science Foundation No 20-71-10048.
\end{acknowledgments}

\section*{Data Availability}

Data sharing is not applicable to this article.


\bibliography{h2dc}

\end{document}